\documentclass[preprint,superscriptaddress]{revtex4-1}

\usepackage{supertabular}
\usepackage{amsmath}
\usepackage{graphicx}
\usepackage[pdftex, colorlinks=true,linkcolor=blue]{hyperref}		

\begin{document}
\title{Guided Lock of a Suspended Optical
  Cavity \\Enhanced by a Higher Order Extrapolation}
\author{Kiwamu Izumi}

\altaffiliation{Current address: LIGO Hanford Observatory, P.O. Box 159 Richland, WA 99354, USA}

\email[Corresponding author: ]{izumi\_k@ligo-wa.caltech.edu}

\affiliation{Department of Astronomy, University of Tokyo, 7-3-1 Hongo, Bunkyo, Tokyo 113-8654, Japan}
\affiliation{National Astronomical Observatory of Japan, 2-21-1 Osawa Mitaka, Tokyo, Japan}

\author{Koji Arai}
\altaffiliation{Current address: LIGO Laboratory, California Institute of Technology, Pasadena, CA 91125, USA}

\affiliation{National Astronomical Observatory of Japan, 2-21-1 Osawa Mitaka, Tokyo, Japan}

\author{Daisuke Tatsumi}
\affiliation{National Astronomical Observatory of Japan, 2-21-1 Osawa Mitaka, Tokyo, Japan}

\author{Ryutaro Takahashi}
\affiliation{National Astronomical Observatory of Japan, 2-21-1 Osawa Mitaka, Tokyo, Japan}

\author{Osamu Miyakawa}
\affiliation{Institute of Cosmic Ray Research, University of Tokyo, 5-1-5 Kashiwanoha, Kashiwa, Chiba, Tokyo, Japan\\}

\author{Masa-Katsu Fujimoto}
\affiliation{National Astronomical Observatory of Japan, 2-21-1 Osawa Mitaka, Tokyo, Japan}
\date{Compiled \today}

\begin{abstract}
{Lock acquisition of a suspended optical cavity can be a highly stochastic process and is therefore nontrivial.
Guided lock is a method to make lock acquisition less stochastic by decelerating the motion of the cavity length based on an extrapolation of the motion from an instantaneous velocity measurement.
We propose an improved scheme which is less susceptible to seismic disturbances by incorporating the acceleration as a higher order correction in the extrapolation. We implemented the new scheme in a 300-m suspended Fabry-Perot cavity and improved the success rate of lock acquisition by a factor of 30.}
\end{abstract}

\maketitle
\section{Introduction}
Suspended optical cavities are ones of the most important building blocks in terrestrial laser interferometric gravitational wave antennae~\cite{Somiya:2012cqg, Abramovici:1992, Bradaschia:1990}.
The suspended optical cavities allow for the intracavity fields to bounce multiple times and therefore increase the interaction time of the laser fields with incoming gravitational waves.
Such an enhancement takes place only when the cavity is in the vicinity of a resonance.
If the frequency of the incident laser is sufficiently stabilized, the only major disturbance is the displacement of the mirrors.
The displacement is typically dominated by seismic vibration which displaces the mirrors by a comparable size to the laser wavelength or 1~$\mu$m.
Therefore, an active control of the cavity length is necessary for maintaining a resonance.

Lock acquisition is a length control process in which a suspended cavity is brought from an uncontrolled state to a controlled state.
When uncontrolled, the cavity passes through resonances in a stochastic manner due to the continuous excitation by seismic vibration.
When the cavity is in the vicinity of a resonance, one can obtain interferometric signals representing the displacement but only in a  narrow range around the resonance (1~nm for cavities with a finesse of a few hundred, typical for gravitational wave antennae).
A naive approach would be to enable a feedback control for the mirror position when the cavity starts to pass through a resonance.
If successful, the cavity stays within the range where the signal is available and therefore one can maintain the resonance by keeping the feedback control.
However, in order to meet the stringent noise requirement, the actuators are typically designed to provide weak force so that the mirrors are less coupled to the electronics.
For this reason, as we design more sensitive gravitational wave antennae, lock acquisition becomes more difficult and can introduce long down time during observing runs and during critical commissioning experiments such as noise hunting~\cite{Matynov:PhysRevD2016}.

To quantitatively illustrate the difficulty, let us think about a simplified model with a mirror with a mass of $m$ moving with an initial velocity $v$.
We shall attempt to stop the mirror by exerting a constant actuation force $F$ longitudinally on the mirror.
The displacement $\Delta L$ until the mirror stops can be derived from the energy conservation law as
	\begin{equation}
  		F\Delta L=\frac{1}{2} m v^2\,.	\label{eq:kinm}
	\end{equation}
For successful lock acquisition, this deceleration process must be completed within the displacement range where an appropriate interferometric signal remains available.
The typical range for such a signal is as large as the full-width-at-half-maximum of the cavity resonance.
Therefore the maximum allowed size for the displacement $\Delta L_\text{max}$ can be given as
	\begin{equation}
  		\Delta L_\text{max}= \frac{\lambda}{2 {\cal F}} \,,
	\end{equation}
where $\lambda$ is the wavelength of the laser and $\cal{F}$ is the finesse of the cavity.
This constrains the maximum mirror velocity, that the actuator can stop, to be
	\begin{equation}
  		v_\text{max}= \sqrt{\frac{\lambda F}{m {\cal F}}}\,. \label{eq:maxv}
	\end{equation}
For instance, this maximum velocity is $2\,\mu\text{m/s}$ for the Japanese prototype gravitational wave antenna, TAMA300~\cite{Arai:2009cqg} with the ideal parameters: $F=1.8\times10^{-3}\,{\rm N}$; $\lambda = 1064 \,{\rm nm}$; $m = 1 \,{\rm kg}$; and ${\cal F} = 500$.
This is comparable to the root-mean-square (rms) of typical measured velocity. This means lock acquisition is a stochastic process which can frequently fail. In fact, we have seen that the success rate was significantly degraded during high seismicity times. Moreover, the effective actuator force can be smaller than the aforementioned value for fast signals due to the frequency response of the electronics and can therefore further deteriorate lock acquisition.

Several approaches have been proposed to make lock acquisition less stochastic~\cite{Mullavey:2012oex,Izumi:2012josaa,Aso:2004pla,Shaddock:2007ol}. An intriguing approach among them is \textit{guided lock} which virtually increases the signal range $\Delta L_\text{max}$ by extrapolating the motion of the cavity~\cite{Camp:1995ol}.
In this approach, the cavity motion is extrapolated from a measured instantaneous velocity at a resonance as the cavity passes through it.
A damping pulse is subsequently applied to a cavity mirror in such a way that the cavity length swings back to the same resonance with a reduced velocity.
Therefore, it increases the success rate of lock acquisition.
Its appealing advantage over the others is that it involves much less hardware preparation; the core hardware is a programmable signal processor. So for this reason, Virgo~\cite{Bradaschia:1990}---one of the gravitational wave antenna projects---recently implemented the guided lock scheme to increase the success rate of lock acquisition for two 3~km  suspended optical cavities~\cite{Bersanetti.PhD}.
However, as Camp {\it et al.}~\cite{Camp:1995ol} reported, a challenge is to make the method robust and reliable against seismic disturbance which deteriorates and occasionally corrupts the extrapolation by stochastically agitating the cavity length.

In this article, we propose an advanced version of the guided lock scheme which is less susceptible to seismic disturbance by incorporating a higher order term of the cavity motion in the extrapolation.
It incorporates the information of the acceleration in addition to the velocity. This consequently maintains the accuracy of the extrapolation and therefore mitigates the corruption caused by seismic disturbance. 
We implemented and tested the scheme in a suspended Fabry-Perot cavity with a length of 300~m in TAMA300. We observed a drastic improvement in lock acquisition. We also discuss limiting factors for the current scheme.

The organization of the paper is given as follows.
In section~\ref{sec:del}, we provide the concept, advantage and requirement of the scheme based on a numerical simulation study.
Section~\ref{sec:imp} describes an implementation of the proposed scheme in a suspended Fabry-Perot cavity at TAMA300.
In section~\ref{sec:decexp}, we show the experimental verification of the deceleration.
Section~\ref{sec:disc} discusses limitations for the deceleration performance.
In section~\ref{sec:success}, we demonstrate an improved success rate of lock acquisition.
Finally, our study is summarized in section~\ref{sec:summary}.

\section{Extrapolation of cavity motion} \label{sec:del}
Guided lock, in general, consists of two distinct operations; an extrapolation of cavity motion and subsequent deceleration based on the extrapolation.
We quantitatively show that the accuracy of the extrapolation can be improved by including a higher order term of the motion.
We then show that in order to achieve a certain accuracy level, the whole guided lock process must finish within a certain duration of time.

\subsection{Polynomial expansion of the cavity trajectory}
To study the accuracy of extrapolations for cavity motion, we shall start from generalizing the motion.
The single trip length of a cavity $x(t)$ can be written as
	\begin{equation}
		x\left( t\right) = L_0 + \sum b_i\xi_i \left(t\right),
	\end{equation}
where $L_0$ is a static single trip length of the cavity, $\xi_i$ is displacement of $i$-th cavity mirror from the equilibrium point, and $b_i$ is a scaler factor representing geometrical effects (e.g., angle of incidence). In the case of a Fabry-Perot cavity, $|b_1|=|b_2|=2$.
In general, the cavity length at a certain time $t_0$ can be expressed by the Taylor series expansion as
	\begin{equation}
		x\left(t \right) = x\left(t_0\right) + \dot{x}\left(t_0\right)\,\left(t-t_0\right) + \frac{1}{2!} \ddot{x}\left( t_0\right)\left(t-t_0 \right)^2+ \ldots. \label{eq:xexp}
	\end{equation}
In guided lock, the parameters are measured at a cavity resonance.
For this reason, we can initialize the time and position i.e., $t_0=0$ and $x(0)=0$ and therefore the first term is irrelevant.

If the mirrors were perfect free masses without external disturbances, the second term---the constant velocity of the cavity---is sufficient to describe the motion in the rest of the time.
In the previous experiment by Camp \textit{et al}~\cite{Camp:1995ol}, the terms up to the constant velocity was incorporated. 
However, in practice, this may not be accurate enough because of the following two reasons. 
Firstly, because the cavity mirrors are suspended as pendulums, the associated restoring forces are present.
Secondly, seismic vibration continuously excites the motion of the mirrors through the suspending wires.
Therefore the true motion does not obey the constant velocity model in reality.
In times of high seismic motion, seismic disturbance can agitate cavity motion so much that it completely spoils the extrapolation and the guided lock technique does not work.

\begin{figure}[t]
	\centerline{\includegraphics[width=0.8\columnwidth ]{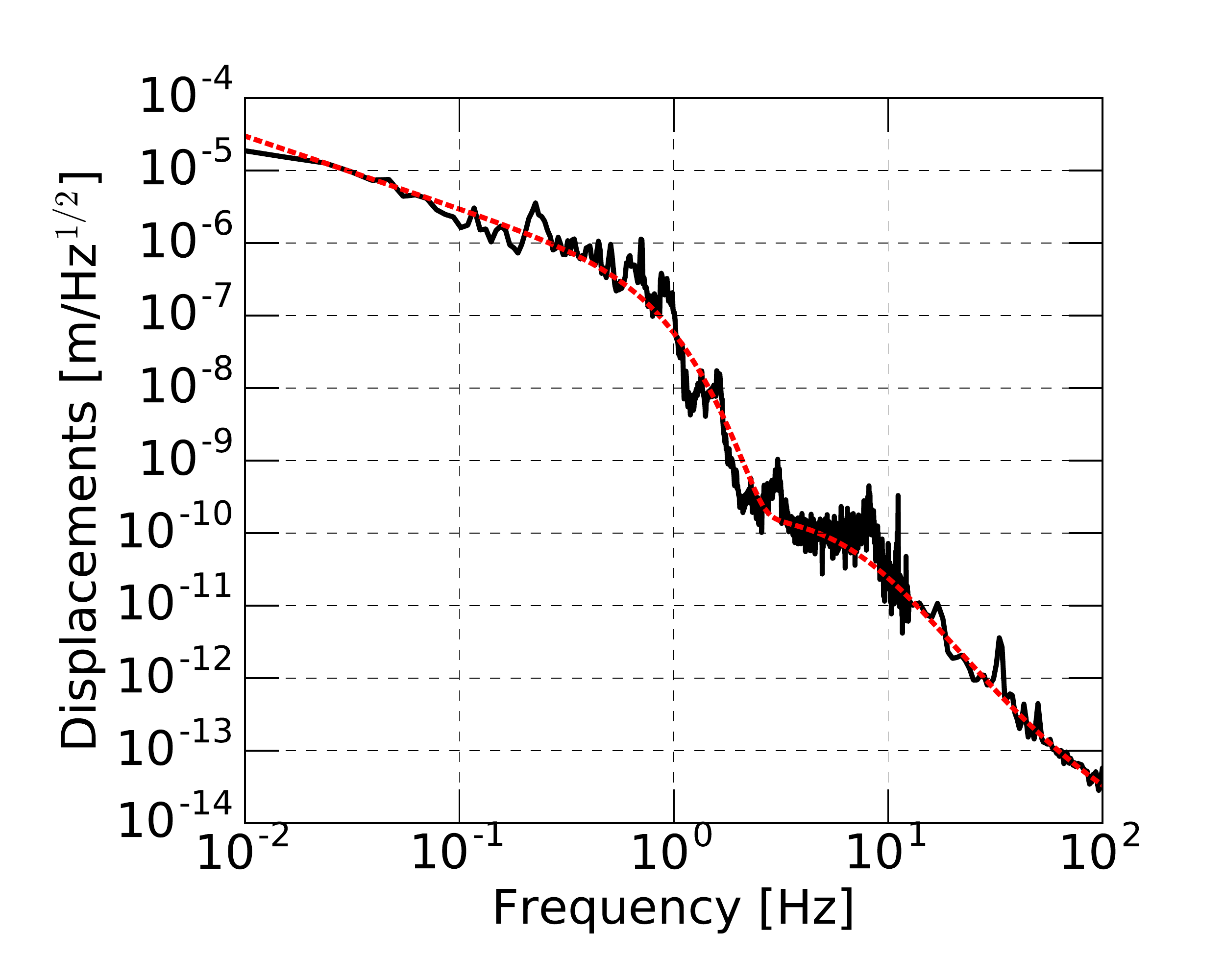}}
	\caption{The simulated displacement of a suspended cavity in amplitude spectral density.
	Solid line: measured length displacement in a 300-m Fabry-Perot cavity at TAMA300.
	Dashed line: modeled displacement from which simulated length fluctuation is generated with zero-mean Gaussian random noise.}
	\label{fig:noise}
\end{figure}

\begin{figure}[t]
	\centerline{\includegraphics[width=0.8\columnwidth ]{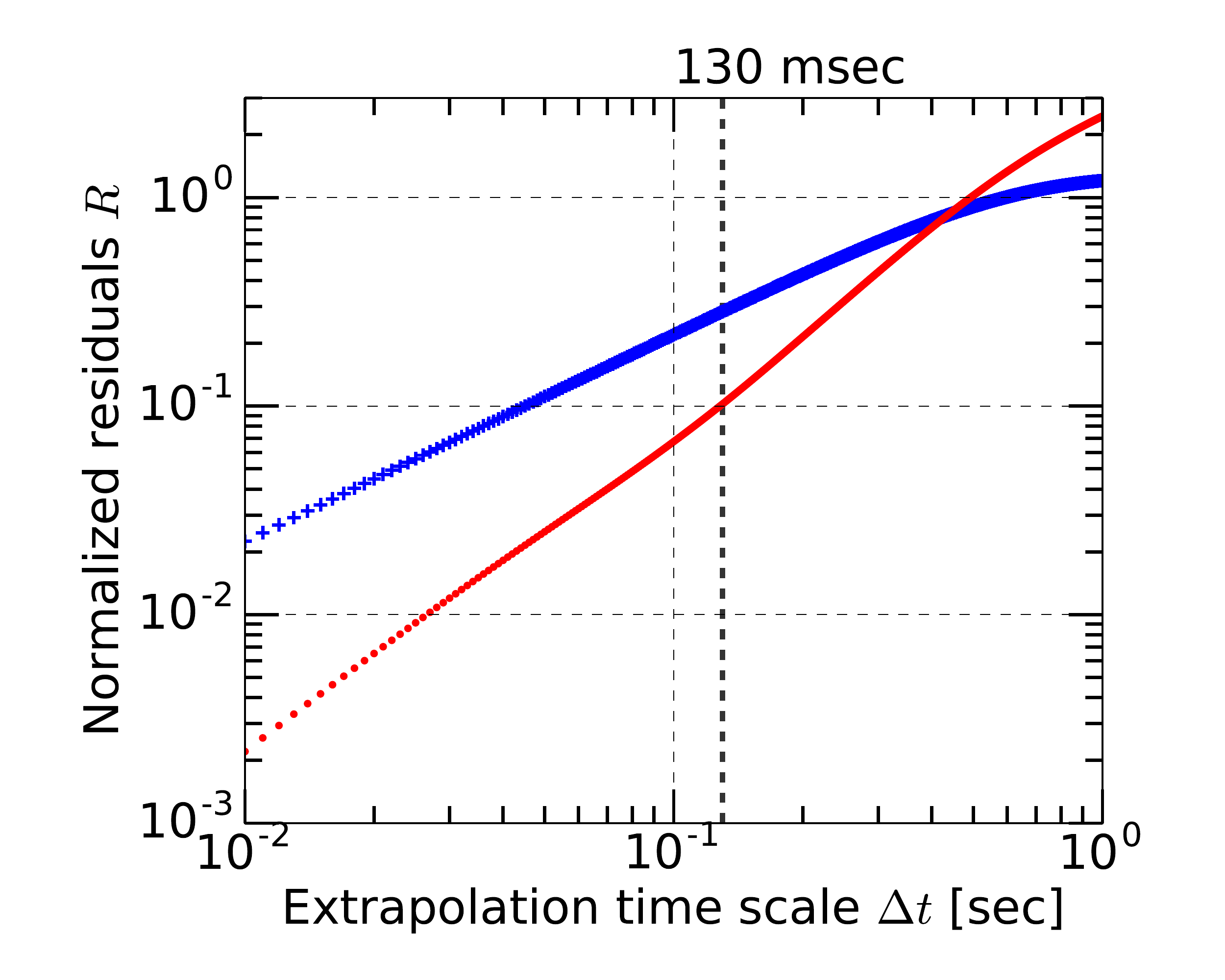}}
	\caption{Normalized residuals $R$ as functions of the extrapolation time scale $\Delta t$.
	Blue crosses: residuals calculated with the constant velocity model $\hat{v_1}$.
	Red dots: residuals calculated with the constant acceleration model $\hat{v_2}$.
}
	\label{fig:residual}
\end{figure}

To improve the accuracy of the extrapolation, we propose incorporating the third term of equation~(\ref{eq:xexp}), the constant acceleration of the cavity motion. 
Even though the new scheme is expected to give an improved accuracy in the extrapolation, it is also obvious that seismic excitation makes even the constant acceleration model inaccurate on a long time scale.
Therefore, it is of high importance to evaluate the accuracy of the extrapolation as a function of the elapsed time.

\subsection{Extrapolation accuracy and its elapsed time}
In order to assess time scales on which the incorporated acceleration improves the accuracy for the extrapolated trajectory, we conducted a numerical simulation.
This assessment in turn places a requirement on the duration of the subsequent deceleration operation so that it finishes before the extrapolation is corrupted by residual seismic disturbance.

We simulated the displacement of a suspended cavity in time series for a duration of 100 sec with a sampling rate of 1~kHz.
The simulated displacement has a colored spectral shape with zero-mean Gaussian distribution as shown in figure~\ref{fig:noise} in order to simulate realistic fluctuations.
We extract the velocity and acceleration $v_0(t_0)$ and $a_0(t_0)$ at a time $t_0$, respectively, which are in turn used as initial parameters for extrapolating the velocity $v (t_0 + \Delta t)$ at a later time after $\Delta t$ elapses.
For comparison, we tested two different extrapolators for the velocity as
	\begin{equation}
	\begin{aligned}
		\hat v_1\left(\Delta t,\,t_0 \right) &= v_0\left(t_0\right),&\\
		\hat v_2\left(\Delta t,\,t_0 \right) &= v_0\left(t_0\right)&+a_0\left(t_0\right) \Delta t. \label{eq:predictors}
	\end{aligned}
	\end{equation}
The first extrapolator $\hat v_1$ uses the constant velocity only, whereas the second one $\hat v_2$ uses both the constant velocity and acceleration.

The accuracy of each extrapolator was then evaluated by taking the residual between the true and extrapolated velocities as
	\begin{equation}
		R_k \left( \Delta t\right)= \left[ \frac{   \sum_{i}^{N}\left[ \hat v_k(\Delta t,\,t_i) - v(t_i + \Delta t) \right]^2 }{ Nv_\text{rms}^2 } \right]^{1/2}\,, \label{eq:residual}
	\end{equation}
where $v_{\rm rms}$ is the rms velocity of the given data in order to normalize the residuals,
$N$ is the total number of evaluations, and subscript $k = (1,2)$ represents the two different extrapolators. 
We slid the starting time $t_i$ from one data point to the next neighboring point all through the data in order to sample as many cases as possible.
The time scale at which the normalized residual becomes unity can be interpreted as a point where the extrapolated velocity is not accurate any more; the size of the residual is as big as that of the spontaneous motion.

Figure~\ref{fig:residual} shows the residuals of the two extrapolators as functions of the elapsed time.
The constant acceleration model, $\hat v_2$, shows smaller residuals than that of the constant velocity model, $\hat v_1$, below 400~msec.
Therefore the constant acceleration model is more accurate than the constant velocity model within this time scale.
If we aim at decelerating the cavity motion by a factor of 10, the residual needs to be suppressed to $10^{-1}$ at least.
This places a requirement on the maximum allowed deceleration time of 130~msec for the extrapolation using the constant acceleration model. The residual of both extrapolators crosses the unity between 500 and 600~msec.
These values can be interpreted as a consequence of the spectral shape of the stochastic motion; because the dominant power of the velocity is concentrated below 1~Hz, the velocity becomes independent of its past history after a fractional cycle of 1~Hz elapses. Thus, a different spectral shape would give a different unity-crossing time. We will discuss the spectral shape of our displacement in the next section.

While the new scheme successfully improves the accuracy for extrapolating the mirror trajectory as the acceleration terms is incorporated, the time scale of the subsequent deceleration process must be within 130 msec in order to fully exploit the improved accuracy.
We describe our implementation in great detail in the next section.

\section{Implementation of the proposed scheme} \label{sec:imp} 
\subsection{Experimental arrangement}
\begin{figure}[t]
	\centerline{\includegraphics[width=0.8\columnwidth ]{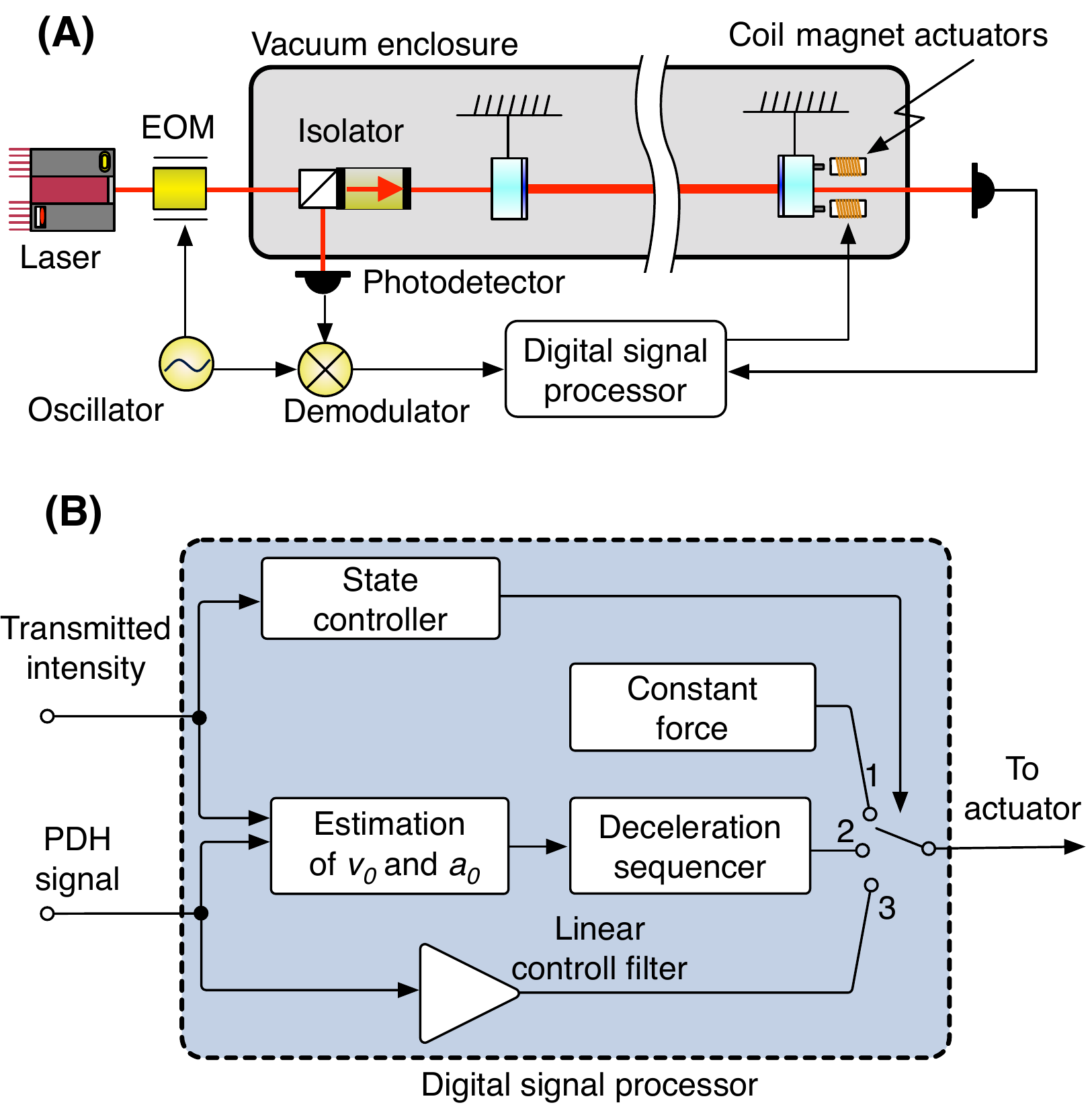}}
	\caption{(A): Schematic view of the experimental setup. EOM stands for electro-optic modulator.
	(B): Block diagram of the signal flow in the digital signal
	processor. The state controller switches the type of the actuation forces depending on what state the guided lock process is.
	The numbers written by the output selector indicate the following types of forces. (1): constant force to push the cavity length back to the same resonance. (2): pre-programmed deceleration forces. (3): linear control signal.
}
	\label{fig:setup}
\end{figure}

A schematic view of the experimental setup is shown in figure~\ref{fig:setup}.
We used a suspended Fabry-Perot cavity with a length of 300~m and a finesse of 500.
This cavity is a part of TAMA300~\cite{Arai:2009cqg} and enclosed in high-vacuum chambers.
The mirrors are suspended by wires, providing a resonant frequency of $1\,{\rm Hz}$.
Seismic vibrations acting on each mirror are attenuated through a multiple stage suspension~\cite{Takahashi:2008cqg}.
With such suspension systems, displacement of a cavity typically exhibits a $1/f^2$ shape below the mechanical resonance frequency of 1~Hz and rolls off steeply above it. Even though the actual spectra can be different depending on the suspension systems, it is generally true that low frequency part of the motion is the predominant component. This characteristic is common for the suspension systems employed in the terrestrial gravitational wave antennae and therefore our proposed guided lock scheme is applicable to those antennae to some extent. 
The position of one of the mirrors can be controlled via a set of coil magnet actuators.

The laser field is generated by a Nd:YAG laser source with the wavelength of $1064$~nm and output power of 10~W.
The frequency of the laser is stabilized using another identical 300-m suspended cavity (not shown in the figure) as a frequency reference~\cite{Nagano:2003RSI}.
The laser field is phase-modulated by an electro-optic modulator at $15$~MHz before the field is sent to the main interferometer.
The reflected light from the cavity is directed to a photodetector whose output is demodulated at the same frequency as that of the modulation, providing the Pound-Drever-Hall readout signal~\cite{Drever:1983apb}.
A digital signal processor (DSP) receives the demodulated signal and subsequently generates a control signal which finally actuates the position of one of the cavity mirrors.
The DSP consists of a 225~MHz clock processing unit, TMS320C6713 from Texas Instruments, and a 16 bit analog-to-digital interface, DSK6713IF/AI2 and AO2 from Hiratsuka Engineering.
The DSP operates at a sampling rate of 200~kHz to achieve the design control bandwidth of 1~kHz for the final control based on a linear control filter.
In addition, the transmitted intensity is detected by an extra photodetector and sent to the same DSP.

\subsection{Algorithm for measuring velocity} \label{sec:vmeas}
To measure the velocity, we use the slope of the Pound-Drever-Hall signal with a correction using the transmitted intensity.
In a quasi-static case where the cavity length varies at a sufficiently slow speed, the slope of the Pound-Drever-Hall signal is sufficient to measure the velocity.
However, in practice, because the cavity can sweep across a resonance before the intracavity field reaches the equilibrium state, the slope of the signal can become shallower~\cite{Izumi:2014ol}.
Moreover, nonlinear distortion of the signals~\cite{Rakhmanov:2001ao} becomes outstanding for velocities larger than $\pi c\lambda/(4L \mathcal{F}^2)$ with $c$ the speed of light. For TAMA, this is approximately $3\,\mu{\rm m/s}$.
These two effects lead to an inaccurate estimate of the velocity if only using the slope of the Pound-Drever-Hall signal.
A numerical, plane-wave, time-domain, interferometer simulator~\cite{e2e:2006} suggests that the absolute value of the velocity would be underestimated by a factor of more than two for a cavity moving at a constant velocity of 4~$\mu$m/s and even more for higher velocities.

To alleviate such a large inaccuracy, we adopted the transmitted intensity as an additional correction term.
We measure the velocity by using the following empirical expression,
	\begin{equation}
		v(t_0)=\frac{1}{H} \frac{d S_\textrm{PDH}}{dt} \bigg| _{t=t_0} \bigg( \frac{T}{T_\textrm{max}}\bigg)^
{-1.4}, \label{eq:gamma}
	\end{equation}
where  $H$ is the optical gain of the Pound-Drever-Hall signal for the cavity on resonance in meters/volts, $S_\textrm{PDH}$ is the Pound-Drever-Hall signal in volts whose time derivative is computed from two data points separated by 0.1~msec at around the time when the signal crosses zero or a resonance, and where $T_\textrm{max}$ and $T$ represent the maximum transmitted intensity (measured beforehand when the cavity is held at a resonance) and the peak height of the transmitted intensity (measured by picking a data point with the highest value when the cavity passes through the resonance), respectively.
The folding number of $-1.4$ for the peak transmittance was empirically chosen from the numerical simulation such that the best linearity is obtained in a wide range of the cavity speed.

This technique provides us with two practical advantages.
Firstly, this scheme is computationally inexpensive.
Because the DSP runs at a relatively high sampling rate to achieve the design control bandwidth, reduction of the computational load is critical for us.
Secondly, the method makes the measurement less sensitive to a change in the optical gain as it is normalized by the transmitted intensity.
However, these advantages come at the cost of small but significant systematic errors in the estimated velocity below 2 $\mu$m/s.
This method overestimates the velocity by a few\% for 2 $\mu$m/s and the size of the overestimation monotonically increases to 10\% for 1$\mu$m/s as the velocity becomes smaller.
We will quantitatively discuss the influence of such remaining systematic errors in section~\ref{sec:disc}.

\begin{figure}[t]
	\centerline{\includegraphics[width=0.5\columnwidth ]{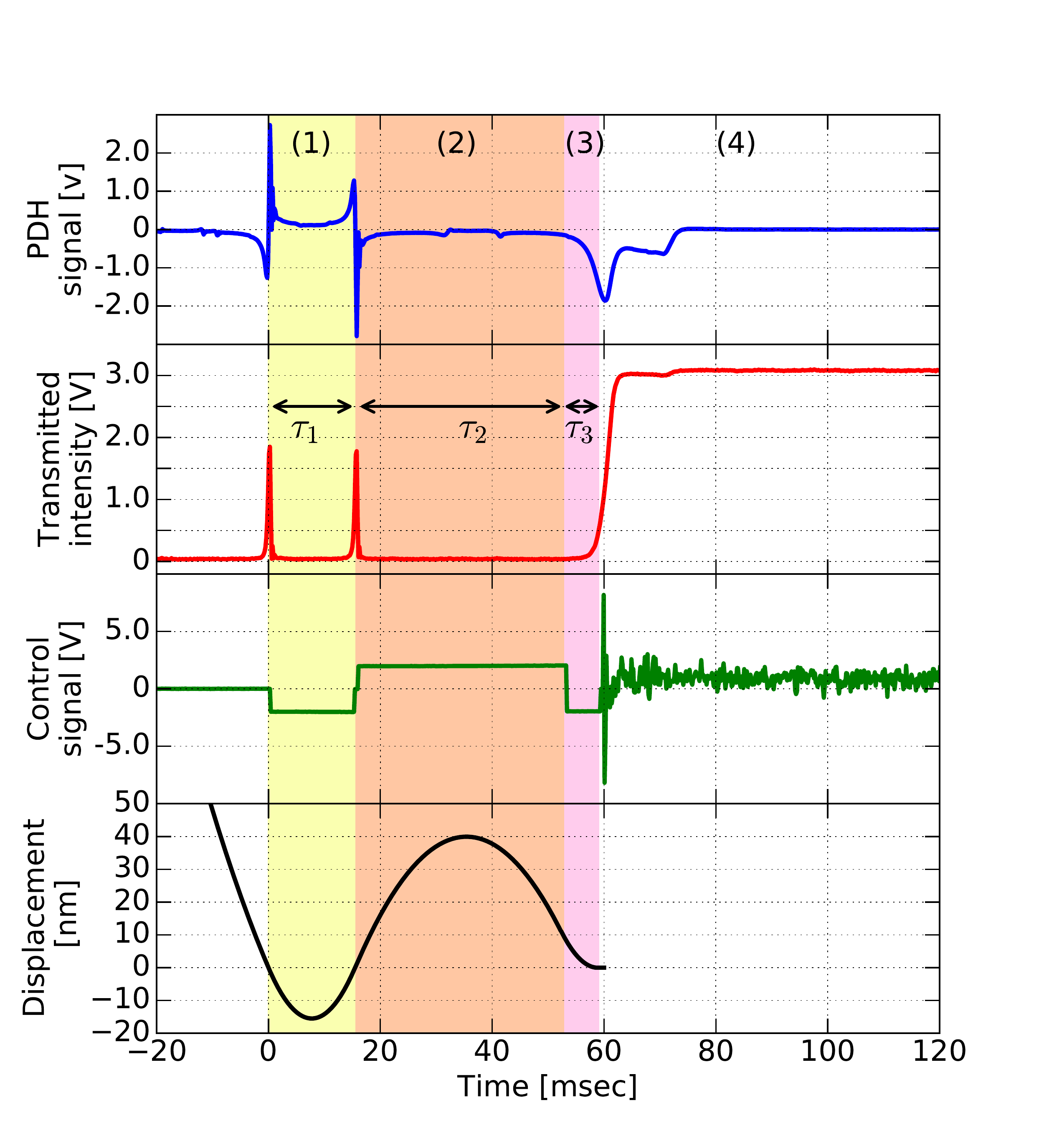}}
	\caption{An actual sequence of the implemented guided lock process in time series.
	The initial parameters are measured to be $v_0 = 4.0$~$\mu$m/s and $a_0=160$~$\mu$m/s$^2$.
	The target velocity ratio is set to $D = 10$\%.
	(Top panel): error signal obtained by the Pound-Drever-Hall technique.
	(2nd top panel): the transmitted intensity,
	(3rd top panel): control signal or equivalently acceleration sent to the cavity mirror. A 2-volt constant control signal corresponds to a constant acceleration of 360~$\mu$m/s$^2$,
	 (Bottom panel): estimated displacement reproduced by a post analysis.
	Annotated colored bands represent the following distinct periods.
	(1): The measurement of the initial conditions by applying a constant force on the mirror to let the cavity length swing back to the same resonance. 
	(2): The first deceleration step, where a constant force is applied for $\tau_2$ to change the direction of the velocity.
	(3): The second deceleration step, where a constant force with the opposite sign is applied for $\tau_3$ to slow down the cavity motion.
	(4): Linear feedback signal is applied to hold the cavity at the resonance.}
	\label{fig:2fringe}
\end{figure} 

\subsection{Algorithm for measuring acceleration}
The acceleration is measured in a different way; we exert a known acceleration to a cavity mirror by using the coil magnet actuator and compare the applied acceleration against the acceleration of the cavity motion.

When the cavity passes a resonance with its instantaneous velocity measured, we apply a constant attractive acceleration $a_\textrm{act}$ to a cavity mirror until the cavity swings back to the same resonance. This operation is annotated as (1) in figure~\ref{fig:2fringe}.

If there was no acceleration, one can precisely predict when the cavity will return to the resonance, based on the initial velocity $v_0$. 
However, because of the presence of the acceleration, the arrival time can be different from that expected without acceleration.
This directly means that one can estimate the acceleration by measuring the time until arrival, $\tau_1$.
We use the following expression to determine the acceleration $a_0$,
	\begin{equation}
		a_0= \frac{2|v_0|}{\tau_1}-|a_\textrm{act}|. \label{eq:tau1}
	\end{equation}
The sign convention for $a_0$ is defined such that a positive value represents an initial acceleration acting to the same direction as the actuator acceleration.
Even though the velocity when the cavity returns to the resonance should be the same as $v_0$ in our model, we measure the velocity again in the second appearance of the resonance and update $v_0$ by the newly measured value.
At this point, both initial velocity and acceleration are in hand; we are ready to decelerate the cavity motion.

 \subsection{Deceleration algorithm}
As discussed in section~\ref{sec:del}, our goal here is to develop a deceleration algorithm which finishes within 130~msec for the cavity moving at the typical largest velocity of 10~$\mu$m/s.

The deceleration algorithm is composed of two steps annotated as (2) and (3) in figure~\ref{fig:2fringe}.
In the first step, a constant force $a_\text{act}$ is applied on a cavity mirror for a duration $\tau_2$ in order to push the cavity length back to the resonance. This operation changes the direction of the velocity.
In the second step, immediately after the first stage, the speed of the cavity is reduced by applying the same amount of constant force with the opposite sign for a duration $\tau_3$ while the cavity length approaches the resonance.
At the time when the second stage finishes, the cavity length arrives back at the resonance with a reduced velocity.
The durations of two steps are predetermined as functions of the initial states $(v_0,a_0)$ as follows.
	\begin{equation}
	\begin{aligned}
		&\tau_2(v_0, a_0) = \frac{|v_0|}{|a_\text{act}| - a_0}\left( M + 1\right), \\
		&\tau_3 (v_0, a_0) = \frac{|v_0|}{|a_\text{act}| + a_0} \left( M - D \right), \\
		\label{eq:tau3}
	\end{aligned}
	\end{equation}
where 
	\begin{equation}
		M =  \left[ \frac{|a_\text{act}| + a_0}{2|a_\text{act}|} + \frac{|a_\text{act}| - a_0}{2 |a_\text{act} |} D^2\right]^{1/2}, \label{eq:M}
	\end{equation}
and where $D$ is the target velocity ratio defined as 
	\begin{equation}
	D=-v_\text{req}/v_0,
	\end{equation}
with $v_\text{req}$ the requested terminal velocity. See \hyperref[sec:appendix]{Appendix} for derivation of $\tau_2$ and $\tau_3$. Since the terminal velocity should be with the opposite sign to the initial velocity, the setting of the velocity ratio is a positive value in the range $0\le D\le 1$.

Because the cavity length wanders back and forth around a resonance during these steps, the whole process takes a duration of approximately $\tau_1+\tau_2+\tau_3 \sim 4v_0/a_\text{act}$ with an additional weak dependence on the acceleration.
Therefore, the entire process can finish in 110~msec for a large initial velocity of $v_0=10$~$\mu$m/s with actuator's acceleration of $a_\text{act} = 360\,\mu{\rm m/s}^2$. We intentionally limit the actuation acceleration to this small value, corresponding to 20\% of the full voltage range, in order not to saturate the actuation electronics.
Nevertheless, it still satisfies our goal time scale of 130~msec.

Now, we evaluate the effect of residual seismic disturbance [i.e., the fourth and higher order terms in equation~(\ref{eq:xexp})]. The effect on the resulting velocity ratio can be computed using the residual given by equation~(\ref{eq:residual}),
	\begin{equation}
	\left|\frac{ \Delta v_\text{d}}{ v_0 } \right| = \frac{v_\text{rms} R_2\left(\tau\right)}{v_0} \quad \text{where} \quad \tau=\sum_{j=1}^3 \tau_i = \frac{4 v_0}{a_\text{act}}, \label{eq:unc}
	\end{equation}
and where $\Delta v_\text{d}$ is a deviation in the terminal velocity from the ideal value.
As shown in figure~\ref{fig:residual}, the residual for our extrapolation, $R_2$, grows roughly as $\tau^{3/2}$ and thus $R_2 \propto v_0^{3/2}$. Plugging this in to the equations above, one can find that the resulting error scales with the initial velocity as $|\Delta v_\text{d} / v_0| \propto v_0^{1/2}$. Therefore the effect of seismic disturbance becomes larger as $v_0$ becomes larger. This is merely due to the fact that a large $v_0$ requires a longer deceleration time. Evaluating this effect, we found that the variation in the resulting velocity ratio can be as high as $\pm4\%$ for a large initial velocity of $10\,\mu{\rm m/s}$. This is smaller than the target velocity ratio of 10\% and therefore prevents seismic disturbance from corrupting the extrapolation as expected. On the contrary, if instead the extrapolation was computed with the constant velocity model, one would obtain a larger variation of $|\Delta v_\text{d} / v_0| \sim 12\%$ almost independently of the initial velocity value. This would lead to corruption events in which the residual seismic disturbance drags the cavity length so hard that the cavity doesn't return to the resonance at around the expected time.

Summarizing this section, we reported a successful implementation of the estimation of the initial state $(v_0, a_0)$ using a DSP and data from the Pound-Drever-Hall and the transmitted power signals. Since the DSP is capable of quickly computing the designed deceleration durations of time based on the measured initial state, it was able to finish the entire process within the required time scale. Applying this scheme, we expect the velocity ratio to be $|v_\text{d}/v_0| =10\pm$a few\% whose deviation is due to residual seismic disturbance.

\begin{table*}[t]
	\begin{center}
	\caption{\bf Summary of the deceleration test.}
		\label{tab:decv}
		\begin{tabular}{c|cccccc}
  			\hline
  			\hline
  			   & $N$ &
			   Avg$(\left| v_0 \right|)$ &
			    Avg$(v_\text{d}/ v_0 )$&
			    Std$(v_\text{d}/ v_0 )$& 
			   Num$\left( |v_\text{d}| < 1 \mu\text{m/s} \right)$ &
			    Num$\left( |v_\text{d}| < 2 \mu\text{m/s} \right)$  \\
  			\hline
			\bf{10\% velocity ratio} & 95 & 4.7~$\mu$m/s & $-0.25$  & 0.18 & 67 (71 \%) & 87 (92\%)\\
			\bf{30\% velocity ratio} & 105 & 4.4~$\mu$m/s &$-0.36$ & 0.12 &  33 (31\%) & 87 (83\%) \\
			\hline
		\end{tabular}
		\caption*{\footnotesize $N$ is the total number of measurements.
		Avg($X$) and Std($X$) represent average and standard deviation of $X$, respectively.
		Num($Y$) is the number of the samples that meet the condition $Y$. The values in the parentheses represent the fractional percentage with respect to the number of data samples.}
	\end{center}
\end{table*}

\section{Experimental verification of deceleration} \label{sec:decexp}

A crucial feature of guided lock is the capability to reduce the cavity motion based on the extrapolation.
We conducted a measurement to evaluate the success rate of the deceleration. The success rate here is defined as the rate of the deceleration attempts satisfying the maximum acquirable speed of $2\,\mu\text{m/s}$ after the application of the proposed deceleration scheme.
In this experiment, the initial parameters $v_0$ and $a_0$ are measured when the cavity length passes through a resonance and subsequently the deceleration is applied.
Then the terminal velocity when the cavity length returned to the resonance was recorded.
At its last passage through the resonance, no linear feedback signal was applied to the mirror to allow measurement of the reduced velocity.
The measurement was repeated approximately 100 times for a particular velocity ratio setting.
We programmed the DSP not to apply a force for 30 sec after every deceleration measurement in order to let the cavity motion settle to the nominal.

The results are summarized in table~\ref{tab:decv}.
Two different velocity ratio settings, $D=10\%$ and $30\%$, were tested.
Ideally, with the 10\% velocity ratio setting and maximum initial velocity of 10~$\mu$m/s, all the samples will be slowed to speeds smaller than $2\,\mu\text{m/s}$, allowing successful lock acquisition.
In contrast, the 30\% velocity ratio will leave a fraction of the samples which end up with a speed higher than $2\,\mu\text{m/s}$.
In the measurement with $D=0.1$, 92\% of the samples had resulting velocities that met the requirement and 71\% of the samples had velocities below $ 1\,\mu\text{m/s}$.
Therefore, the measured success rate was indeed high, but didn't reach our expectation of 100\%.
For $D=0.3$, 83\% of the samples had resulting velocities that met the requirement and 31\% had velocities below $1\,\mu\text{m/s}$.
From these results, we can conclude that the 10\% setting more reliably meets the requirement.

When the velocity ratio was set to 30\%, the terminal velocities showed a good agreement with the expected values.
The ratio of final to initial velocities was measured to be 36\% on average.
On the other hand, when the ratio was set to 10\%, the terminal velocities significantly deviated from the expected values.
The average ratio was measured to be 25\% which is a factor of 2.5 larger than it should be. Moreover, the standard deviation of the velocity ratio was measured to be 18\% which is significantly larger than the expected value from residual seismic disturbance. In the next section, we will discuss these results by studying possible limiting factors for the deceleration performance.


\section{Discussion on the deceleration} \label{sec:disc}
In practice, the performance of the deceleration can be influenced by errors in estimating the initial state ($v_0, a_0$).
Specifically, we found errors in the initial velocity critical for achieving the desired velocity ratio. We introduce a fractional error $\delta$ in the estimated velocity so that the estimated velocity is expressed by $v_0(1+\delta)$.
A negative error represents an underestimated initial velocity, whereas a positive error represents an overestimation.
For a small error ($|\delta| \ll 1$), the resulting ratio $|v_\text{d}/v_0|$, can be analytically obtained as
	\begin{equation}
		\left| \frac{v_\text{d}}{v_0} \right|= \begin{cases} D + \delta \left(1+D\right),&  \left( \text{for }\delta <0 \right),\\
				D + \delta \left(D^{-1} +D + \sqrt{2 + 2D^{-2}}  \right), &  \left(\text{for } \delta >0 \right),\end{cases}  \label{eq:dec_positived}
	\end{equation}
where we assumed the initial acceleration to be zero ($a_0 =0$) for simplicity. See \hyperref[sec:appendix]{Appendix} for derivation.

As shown in the above expressions, the resulting velocity ratio is a function of both the measurement error and the requested velocity ratio. In the case of an underestimated initial velocity, it results in a terminal speed slower than that expected. For an aggressive deceleration or a small $D$, the deviation in the resulting velocity ratio is almost independent of the requested ratio as $|v_\text{d} /v_0| \approx D+\delta$. While a few \% measurement error doesn't pose significant issue for achieving $D=10$~\% in this case, the situation is completely different for the overestimated initial velocities. An overestimated initial velocity causes an insufficient deceleration, making the terminal speed larger than that expected. In this case, the size of the deviation in the resulting velocity ratio is a strong function of the ratio setting as $|v_\text{d}/v_0| \approx D+ (1+\sqrt{2})\delta/D$ for a small $D$. This difference arises from the fact that, in the overestimation case, the cavity length returns to the resonance before the slowing process $\tau_3$ completes. This consequently results in a larger deviation. Since these two distinct behaviors take place stochastically, we expect the deviations in the resulting velocity ratio to be asymmetrical around the median value.

\begin{figure}[t]
	\centerline{\includegraphics[width=0.8\columnwidth ]{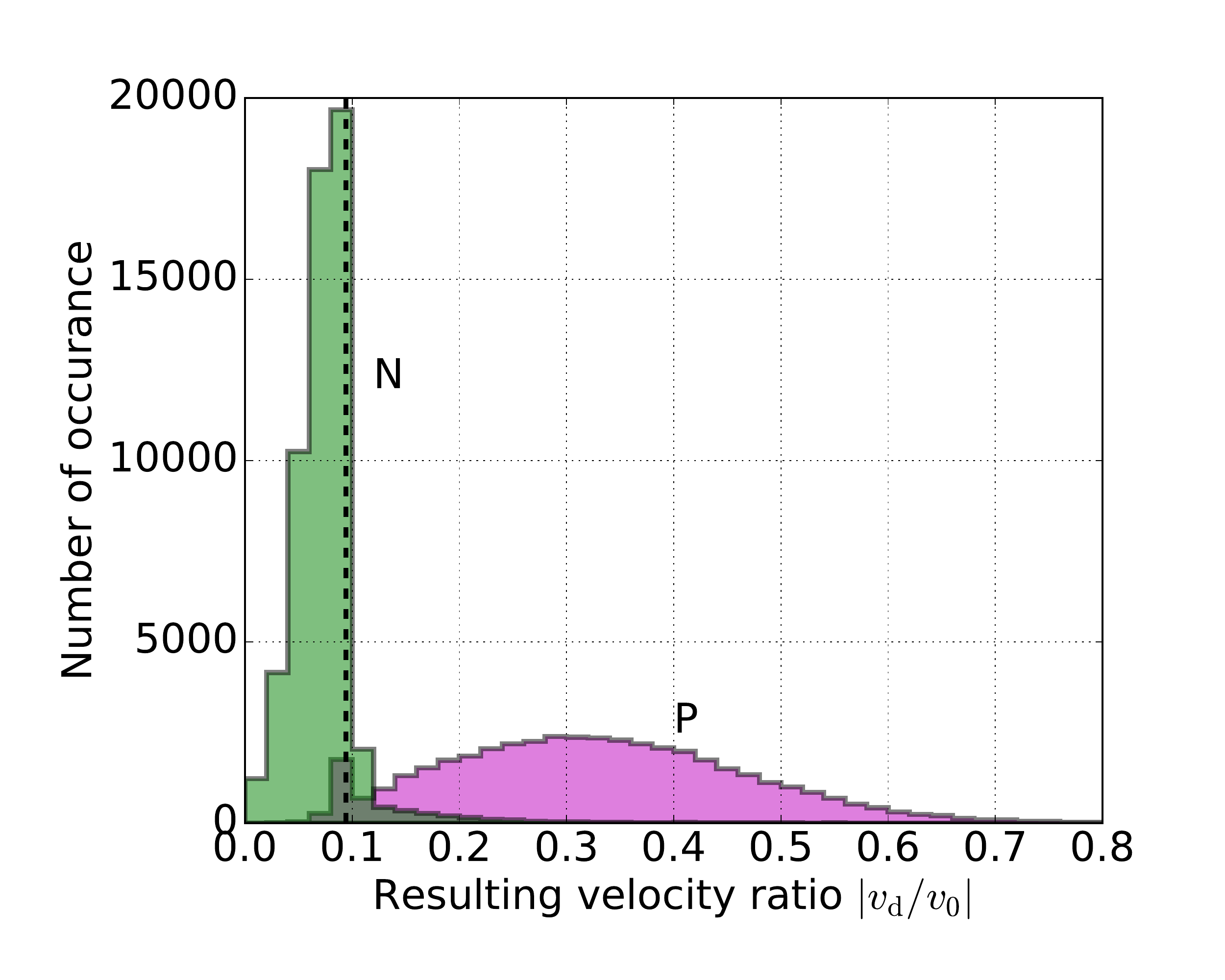}}
	\caption{Distribution of the simulated velocity reductions for a fixed initial velocity of $5\,\mu{\rm m/s}$. 
	(N): the distribution of the resulting velocity ratio influenced by negative error or $\delta<0$. 
	(P): the distribution of the resulting velocity ratio influenced by positive error or $\delta>0$. 
	Regions where the two distribution overlap are shown with a darker color.
	The dashed line represents the median values of the entire distribution.}
	\label{fig:2n5}
\end{figure}

To thoroughly evaluate the influence of measurement errors including those in the initial velocity as well as in the initial acceleration, we conducted a Monte Carlo simulation where $10^5$ sets of randomly sampled initial accelerations, and fractional errors in the initial velocity and acceleration measurements are generated for a fixed true initial velocity.
The fourth and higher order terms $\mathcal O (t^3)$ of the cavity motion in equation~(\ref{eq:xexp}) are not simulated for simplicity. The acceleration was drawn from a zero-mean Gaussian distribution with a standard deviation of $36\,\mu\text{m/s}^2$, the typical rms in our experiments.
Similarly, the fractional errors in the estimated initial velocity and estimated acceleration are drawn from two independent zero-mean Gaussian distributions with a standard deviation of 3.0\% and 4.2\% respectively to simulate the realistic statistical errors.
The known systematic error in our velocity estimation, as mentioned in section~\ref{sec:imp}, was also incorporated in the estimated initial velocities.

Figure~\ref{fig:2n5} shows the variation of the resulting velocity ratio due to the estimation errors when the true initial velocity is set to $5\, \mu{\rm m/s}$. The target ratio was set to $D=10$\%.
We first confirmed that errors in the initial acceleration measurement didn't appreciably change the distribution of the resulting ratio by running the simulation with and without the random errors in the acceleration measurement while keeping the random errors in the initial velocity. This means that the errors in the initial velocity measurement are the primary cause of the resulting distribution. As expected from the analytical argument, the distribution of the velocity ratio was asymmetrical and in fact it tends to form a bimodal distribution.  The sharp distribution at small velocity ratios was found to be due to the underestimation ($\delta <0$) whereas the broad distribution at large velocity ratios was due to the overestimation ($\delta >0$). 

\begin{figure}[t]
	\centerline{\includegraphics[width=0.8\columnwidth ]{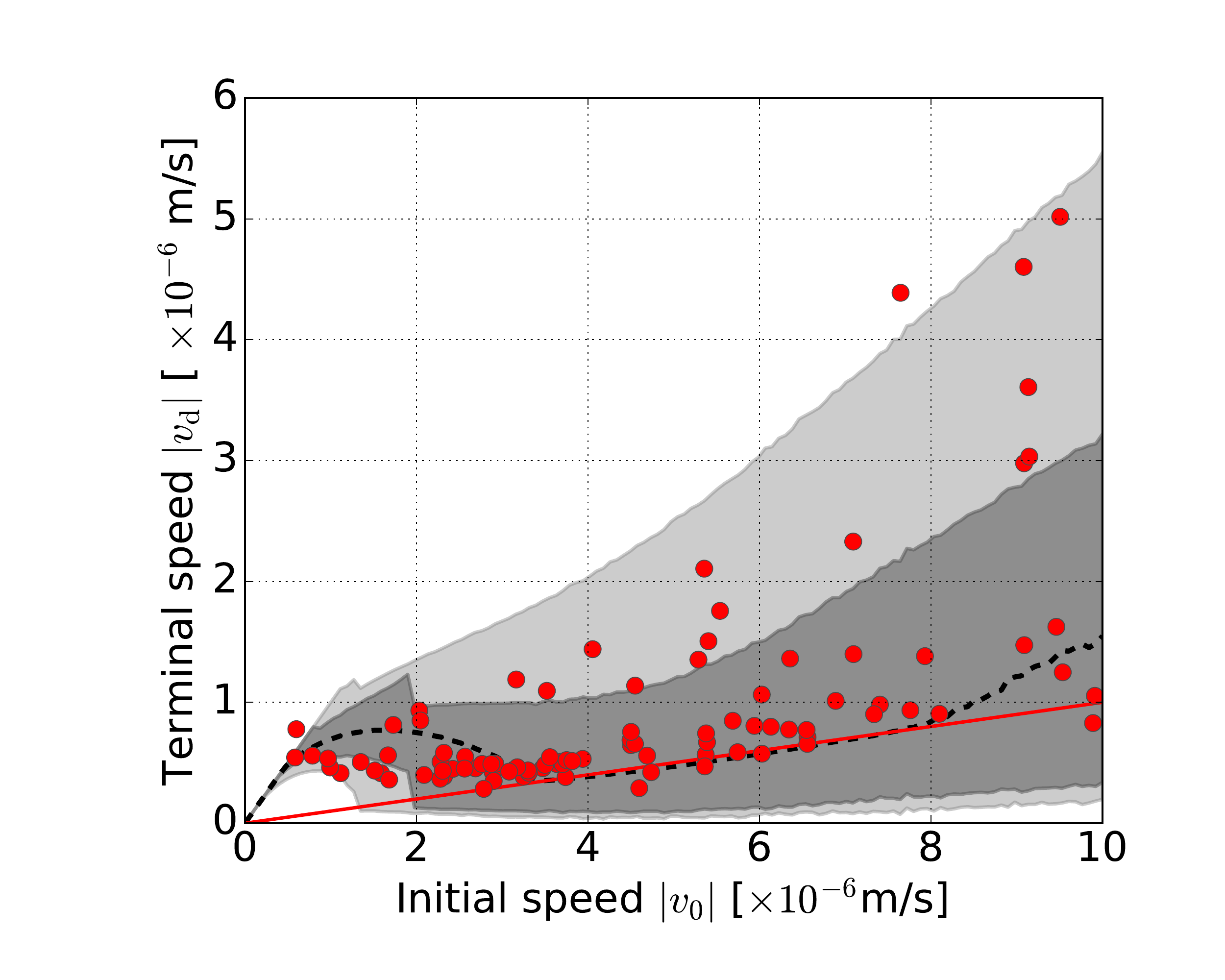}}
	\caption{Comparison of the simulated terminal velocities and those obtained by the measurement for the 10\% velocity ratio setting.
	Dashed line: the median values of the resulting speeds obtained by the Monte-Carlo simulation.
	Dark shaded area: 68\% interval obtained by the simulation.
	Lightly shaded area: 95\% interval obtained by the simulation.	
	Dots: the measured values from the experiments described in section~\ref{sec:decexp}.
	Solid line: the ideal values for the 10\% velocity ratio setting.}
	\label{fig:errorb}
\end{figure}

Finally, the simulation was repeated for different initial velocities ranging from 0.01 to $10\,\mu\text{m/s}$.
A comparison between the simulation and actual measurement is shown in figure~\ref{fig:errorb}.
The simulation shows a good agreement with the measured data.
The median values of the simulated terminal velocity (dashed line in the figure) were found to be precisely due to the remaining systematic error in the velocity measurement (section~\ref{sec:imp}).
The systematic error introduced a relatively large bias in the resulting terminal velocity for the samples with the initial velocities smaller than $2\,\mu{\rm m/s}$. The broad distribution tail towards larger terminal velocities is consistent with the measurement.
We also confirmed that the measured terminal velocities for $D=30\%$ showed a good agreement with the simulation too. These results suggest that the performance of the deceleration is limited largely by errors in estimating the initial velocity.

\section{Success rate of lock acquisition} \label{sec:success}
To evaluate how reliably the scheme can acquire resonance, we measured the success rate of lock acquisition with the guided lock scheme applied. 
In contrast to the deceleration experiment, the DSP starts a linear feedback control as soon as the cavity returns to the resonance regardless of whether the deceleration process has completed or not.
If the linear controller maintains the resonance, this trial is recorded as a success.
The criteria of successful acquisition is given by the transmitted intensity whose value must be more than 20\% of its maximum for $100\,{\rm msec}$ continuously.
If it fails, the DSP is programmed to make another attempt at the guided lock process for the next resonance and repeat the processes until it succeeds.
The lock trial was performed until 100 locks were acquired either with or without the guided lock. The 10\% velocity ratio was used when guided lock was applied. After a successful acquisition, the DSP releases the control by completely shutting off the output force and waiting for 30 sec to let the cavity motion settle down. The number of attempts until solid lock is acquired were recorded for each lock.

%
 
\begin{figure}[t]
	\centerline{\includegraphics[width=0.8\columnwidth]{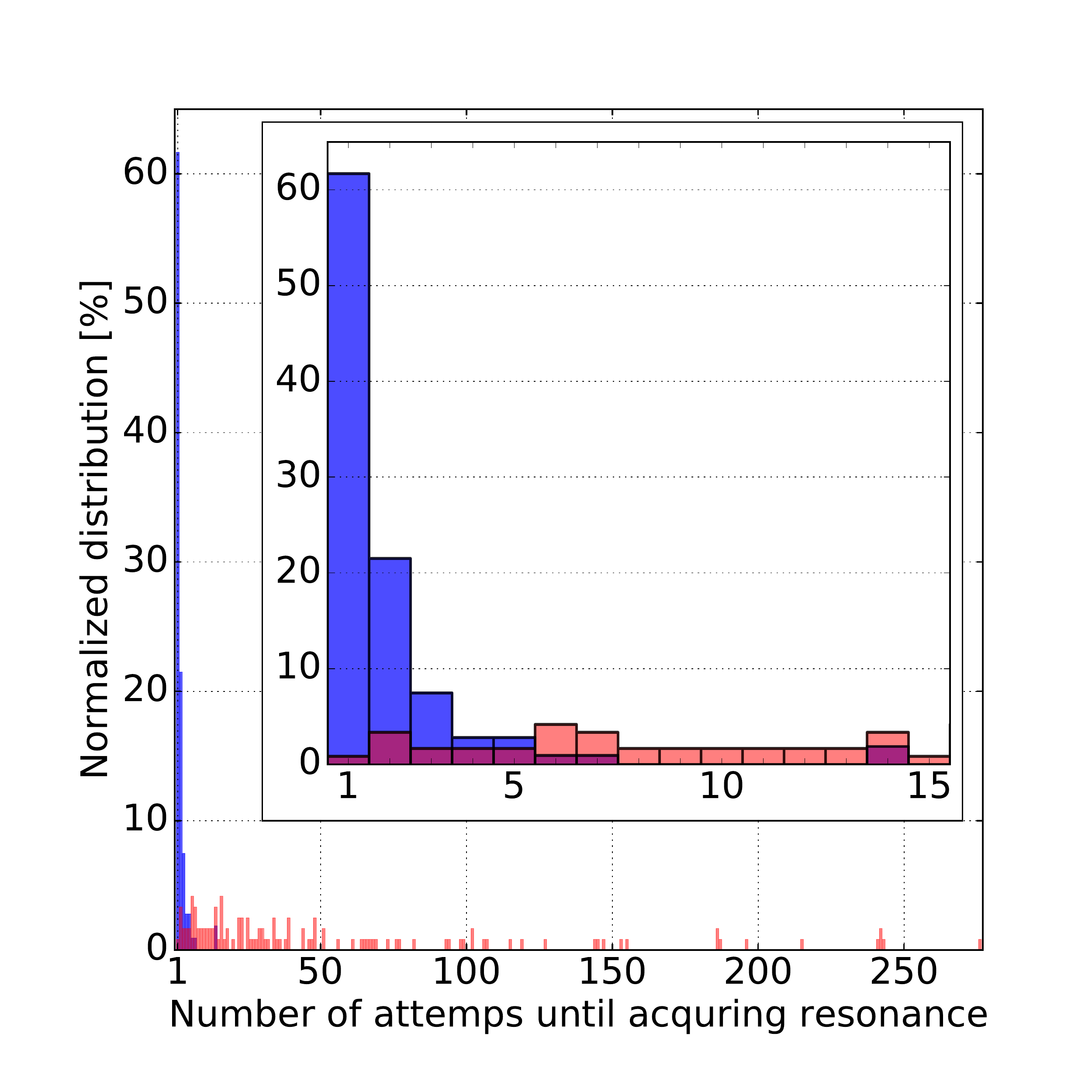}}
	\caption{Measured distribution of the number of attempts until a resonance is acquired.
	Blue colums: the measured distribution with the guided lock scheme applied.
	Red colums: the measured distribution without the guided lock.
	Regions where the two plots overlap are shown with a darker color. 
	The inset shows a zoomed version of the same plot.}
	\label{fig:lockP}
\end{figure}

Figure~\ref{fig:lockP} shows how the guided lock scheme helps lock acquisition.
We observed a drastic improvement in the success rate of lock acquisition by using the guided lock scheme.
The measurement with the guided lock scheme acquired a resonance on the first attempt 62\% of the trials, whereas without guided lock the success rate on the first attempt was only 2\%.
On average, the number of attempts until acquiring a resonance was measured to be 1.6 for guided lock whereas it was 48 without guided lock.
This corresponds to a factor of 30 improvement in the acquisition rate.

\section{ Summary} \label{sec:summary}
We proposed and experimentally tested an advanced guided lock scheme at a 300-m suspended Fabry-Perot cavity in TAMA300. The scheme incorporates not only the initial velocity but also the initial acceleration as a higher order correction for extrapolating the mirror trajectory.
The addition of the acceleration term makes the scheme less susceptible to seismic disturbance and thus more reliably satisfies the speed requirement.
We implemented a two-step deceleration algorithm which finishes the entire process within the required time scale of 130 msec in order to achieve a factor of 10 deceleration.
The new guided lock scheme acquired a resonance with a success rate of 62\% on the first attempt.
This corresponds to a factor of 30 improvement in the average number of attempts, comparing to the one without guided lock.
While the scheme successfully mitigated the corruption of the extrapolation due to residual seismic disturbance, we found that errors in the initial velocity measurement can deteriorate the deceleration performance.
Given the fact that such errors are not fundamental obstacles, we conclude that guided lock can be a powerful scheme toward achieving deterministic lock acquisition of a suspended optical cavity.

\section*{Acknowledgment}
The authors greatly acknowledge T.~Akutsu and S.~Dwyer for fruitful discussion, and H.~Yamamoto for providing the e2e simulation codes.

\section*{Funding}
This work was supported in part by the grant-in-aid for Scientific Research of the Ministry of Education, Culture, Sports, Science and Technology in Creative Scientific Research (09NP0801) and in Priority Areas (415). Development of TAMA-SAS was supported by the Advanced Technology Center of National Astronomical Observatory of Japan and the US National Science Foundation under the cooperative agreement no. PHY-0107417.

\appendix
\section*{Appendix: derivations}
\label{sec:appendix}
Let us start from deriving $\tau_1$, the time for the cavity length to return to the same resonance with a constant actuation force applied.
The cavity length $x$ and its velocity $v$ can be expressed as functions of time,
	\begin{eqnarray}
		x(t) &=& \left| v_0 \right| t - \frac{1}{2} a t^2, \label{eq:app1}\\
		v(t) &= &\left| v_0\right| - a t \label{eq:app2}
	\end{eqnarray}
where we have initialized the length and time so that $x = 0$ when $t=0$, and where $a$ is the effective acceleration given by the sum of two forces,
	\begin{equation}
		a = \left| a_\text{act} \right|+ a_0. \label{eq:a}
	\end{equation}
As we took the absolute value of the initial velocity and actuator force, the cavity length $x$ is defined such that it always departs from the resonance for positive values while the actuator pushes the length back towards the resonance. 
The effective acceleration is increased (decreased) when the initial acceleration is $a_0\ge 0$ ($a_0 \le 0$).
Setting the left hand side of equation~(\ref{eq:app2}) to $-\left| v_0 \right|$ and plugging equation~(\ref{eq:a}), one can obtain
	\begin{equation}
		\tau_1 = \frac{2 \left| v_0\right|}{\left| a_\text{act}\right| +a_0},
	\end{equation}
which is equivalent to equation~(\ref{eq:tau1}).

Before deriving the deceleration durations $\tau_2$ and $\tau_3$, we combine equations~(\ref{eq:app1}) and (\ref{eq:app2}) and get rid of time $t$, so that
	\begin{equation}
		x = \frac{  v_0^2 - v^2}{2a}. \label{eq:xtra}
	\end{equation}
This equation describes the trajectory of the cavity motion in $x$-$v$ plane.
The trajectory during $\tau_2$ can be expressed by substituting $a=\left| a_\text{act}\right| -a_0$ in the above equation. The sign of the initial acceleration is changed because the direction of the actuation force is opposite from that for $\tau_1$. Once $\tau_2$ elapses, the direction of the actuation force is flipped again, making the trajectory 
	\begin{equation}
		x' = \frac{v'^2 - v^2_\text{req} }{\left| a_\text{act}\right| + a_0}, \label{eq:xtra2}
	\end{equation}
where $v_\text{req}$ is the requested terminal velocity given by $D v_0 = -v_\text{req}$.
As we aim at making the trajectory transition from (\ref{eq:xtra}) to (\ref{eq:xtra2}), 
we can derive the velocity at the transition point $v_\times$ by setting $x=x'$ using equations (\ref{eq:xtra}) and (\ref{eq:xtra2}) as
	\begin{equation}
		v_\times = - \left| v_0\right| M,  \label{eq:vcross}
	\end{equation}
where the definition of $M$ is given in equation~(\ref{eq:M}).
Using the simple velocity equation~(\ref{eq:app2}), one can build an equation for $\tau_2$ as
	\begin{equation}
		v_\times  = \left| v_0 \right| - \left( \left| a_\text{act}\right| - a_0\right) \tau_2.
	\end{equation}
Plugging equation~(\ref{eq:vcross}) in to the above, one can obtain $\tau_2$ as expressed in equations~(\ref{eq:tau3}).
Similarly, the velocity must be slowed from $v_\times$ to $v_\text{req}$ during $\tau_3$, so that
	\begin{equation}
		v_\text{req} = v_\times + \left( \left|a_\text{a}  \right| + a_0\right) \tau_3.
	\end{equation}
Plugging equation~(\ref{eq:vcross}) yields $\tau_3$ as expressed in equations~(\ref{eq:tau3}).	
	
Now, let us propagate small errors in the initial velocity. The initial velocity is given by $v_0\left( 1 + \delta \right)$ where $\delta$ represents a small fractional error. This causes incorrect actuation durations which, from equations~(\ref{eq:tau3}), can be expressed by
	\begin{equation}
		\tau'_2 = \tau_2 \left( 1+ \delta \right) 	\quad\text{and}\quad 	\tau'_3 = \tau_3 \left( 1+ \delta \right). \label{eq:taus}
	\end{equation}
Plugging $\tau'_2$ into equations~(\ref{eq:app1}) and (\ref{eq:app2}), one can calculate the cavity length and velocity at the point where the transition of the trajectory takes place 
	\begin{eqnarray}
		x'_\times &=& \left| v_0 \right| \tau'_2  -\frac{\left| a_\text{act} \right|}{2 } \left( \tau'_2\right)^2, \label{eq:dxc}\\
		v'_\times &=& \left| v_0 \right|-\left| a_\text{act} \right| \tau'_2. \label{eq:dvc}
	\end{eqnarray}
Similarly, the cavity length and velocity at the end of the $\tau_3$ actuation can be given by
	\begin{eqnarray}
		x'_\text{d} &=& x'_\times + v'_\times \tau'_3+\frac{\left| a_\text{act} \right|}{2 } \left( \tau'_3\right)^2, \label{eq:dxd}\\
		v'_\text{d} &=& v'_\times +\left| a_\text{act} \right| \tau'_3,
	\end{eqnarray}
Plugging equations~(\ref{eq:tau3}), (\ref{eq:taus}), (\ref{eq:dxc}) and (\ref{eq:dvc}) into the above and setting $a_0=0$, one can obtain the terminal length and velocity
	\begin{eqnarray}
		x'_\text{d} 	&=& -\delta \frac{1-D + \sqrt{2 + 2D^2}}{\left| a_\text{act} \right|} v^2_0 + \mathcal{O}\left( \delta^2\right),\\
		v'_\text{d} &=& -\left| v_0 \right| \left[ D + \delta \left( 1 + D\right)\right].
	\end{eqnarray}
Therefore, the cavity length is not exactly back at the resonance when the $\tau_3$ actuation completes. For the case of underestimated velocities or $\delta <0$, the terminal length $x'_\text{d}$ ends up with a value greater than zero, indicating that the cavity didn't return to the resonance yet. As long as the estimation error is small enough (i.e. $-D/\left(1+D\right) \leq \delta < 0$), the cavity length can then cruise back to the resonance with the constant velocity $v'_\text{d}$.

In contrast, an overestimated velocity or $\delta >0$ ends up with a terminal length smaller than zero, indicating that the the cavity already came back to and passed through the resonance before finishing the $\tau_3$ actuation. In the actual measurements, the velocity is measured when the cavity length is passing through the resonance.
Setting $x'_\text{d} = 0$ in equation~(\ref{eq:dxd}) and replacing $\tau_3'$ with $T$ and solving it for $T$, one can obtain the time for the cavity to return to the resonance from $x'_\times$,
	\begin{equation}
		T = \frac{ -v'_\times -\sqrt{ \left( v'_\times\right)^2 - 2 \left| a_\text{act} \right|  x'_\times } }{\left| a_\text{act}\right|}.
	\end{equation}
Therefore, the velocity at the time the cavity is passing through the resonance can be calculated as
	\begin{equation}
	\begin{split}
		v''_\text{d} &= v'_\times + \left| a_\text{act} \right| T \\
		& = -D \left| v_0 \right|- \frac{1 + D^2 + \sqrt{2 + 2 D^2} }{D} \left|v_0 \right| \delta  + \mathcal{O}\left( \delta^2\right).
	\end{split}
	\end{equation}

\bibliography{references_abb}{}

\begin{thebibliography}{10}
\newcommand{\enquote}[1]{``#1''}

\bibitem{Somiya:2012cqg}
K.~Somiya, \enquote{Detector configuration of {KAGRA}--the japanese cryogenic
  gravitational-wave detector,} {Class. and Quantum Grav.} \textbf{29}, 124007
  (2012).

\bibitem{Abramovici:1992}
A.~Abramovici, W.~E. Althouse, R.~W.~P. Drever, Y.~G{\"u}rsel, S.~Kawamura,
  F.~J. Raab, D.~Shoemaker, L.~Sievers, R.~E. Spero, K.~S. Thorne, R.~E. Vogt,
  R.~Weiss, S.~E. Whitcomb, and M.~E. Zucker, \enquote{{LIGO}: The laser
  interferometer gravitational-wave observatory,} Science \textbf{256},
  325--333 (1992).

\bibitem{Bradaschia:1990}
C.~Bradaschia, R.~D. Fabbro, A.~D. Virgilio, A.~Giazotto, H.~Kautzky,
  V.~Montelatici, D.~Passuello, A.~Brillet, O.~Cregut, P.~Hello, C.~Man,
  P.~Manh, A.~Marraud, D.~Shoemaker, J.~Vinet, F.~Barone, L.~D. Fiore,
  L.~Milano, G.~Russo, J.~Aguirregabiria, H.~Bel, J.~Duruisseau, G.~L. Denmat,
  P.~Tourrenc, M.~Capozzi, M.~Longo, M.~Lops, I.~Pinto, G.~Rotoli, T.~Damour,
  S.~Bonazzola, J.~Marck, Y.~Gourghoulon, L.~Holloway, F.~Fuligni, V.~Iafolla,
  and G.~Natale, \enquote{The {VIRGO} project: A wide band antenna for
  gravitational wave detection,} {Nucl. Instrum. Methods Phys. Res., Sect. A}
  \textbf{289}, 518 -- 525 (1990).

\bibitem{Matynov:PhysRevD2016}
D.~V. Martynov, E.~D. Hall, B.~P. Abbott, R.~Abbott, T.~D. Abbott, C.~Adams,
  R.~X. Adhikari, R.~A. Anderson, S.~B. Anderson, K.~Arai, M.~A. Arain, S.~M.
  Aston, L.~Austin, S.~W. Ballmer, M.~Barbet, D.~Barker, B.~Barr, L.~Barsotti,
  J.~Bartlett, M.~A. Barton, I.~Bartos, J.~C. Batch, A.~S. Bell, I.~Belopolski,
  J.~Bergman, J.~Betzwieser, G.~Billingsley, J.~Birch, S.~Biscans, C.~Biwer,
  E.~Black, C.~D. Blair, C.~Bogan, R.~Bork, D.~O. Bridges, A.~F. Brooks,
  C.~Celerier, G.~Ciani, F.~Clara, D.~Cook, S.~T. Countryman, M.~J. Cowart,
  D.~C. Coyne, A.~Cumming, L.~Cunningham, M.~Damjanic, R.~Dannenberg,
  K.~Danzmann, C.~F. D.~S. Costa, E.~J. Daw, D.~DeBra, R.~T. DeRosa,
  R.~DeSalvo, K.~L. Dooley, S.~Doravari, J.~C. Driggers, S.~E. Dwyer,
  A.~Effler, T.~Etzel, M.~Evans, T.~M. Evans, M.~Factourovich, H.~Fair,
  D.~Feldbaum, R.~P. Fisher, S.~Foley, M.~Frede, P.~Fritschel, V.~V. Frolov,
  P.~Fulda, M.~Fyffe, V.~Galdi, J.~A. Giaime, K.~D. Giardina, J.~R. Gleason,
  R.~Goetz, S.~Gras, C.~Gray, R.~J.~S. Greenhalgh, H.~Grote, C.~J. Guido, K.~E.
  Gushwa, E.~K. Gustafson, R.~Gustafson, G.~Hammond, J.~Hanks, J.~Hanson,
  T.~Hardwick, G.~M. Harry, J.~Heefner, M.~C. Heintze, A.~W. Heptonstall,
  D.~Hoak, J.~Hough, A.~Ivanov, K.~Izumi, M.~Jacobson, E.~James, R.~Jones,
  S.~Kandhasamy, S.~Karki, M.~Kasprzack, S.~Kaufer, K.~Kawabe, W.~Kells,
  N.~Kijbunchoo, E.~J. King, P.~J. King, D.~L. Kinzel, J.~S. Kissel,
  K.~Kokeyama, W.~Z. Korth, G.~Kuehn, P.~Kwee, M.~Landry, B.~Lantz, A.~Le~Roux,
  B.~M. Levine, J.~B. Lewis, V.~Lhuillier, N.~A. Lockerbie, M.~Lormand, M.~J.
  Lubinski, A.~P. Lundgren, T.~MacDonald, M.~MacInnis, D.~M. Macleod,
  M.~Mageswaran, K.~Mailand, S.~M\'arka, Z.~M\'arka, A.~S. Markosyan, E.~Maros,
  I.~W. Martin, R.~M. Martin, J.~N. Marx, K.~Mason, T.~J. Massinger,
  F.~Matichard, N.~Mavalvala, R.~McCarthy, D.~E. McClelland, S.~McCormick,
  G.~McIntyre, J.~McIver, E.~L. Merilh, M.~S. Meyer, P.~M. Meyers, J.~Miller,
  R.~Mittleman, G.~Moreno, C.~L. Mueller, G.~Mueller, A.~Mullavey, J.~Munch,
  L.~K. Nuttall, J.~Oberling, J.~O'Dell, P.~Oppermann, R.~J. Oram, B.~O'Reilly,
  C.~Osthelder, D.~J. Ottaway, H.~Overmier, J.~R. Palamos, H.~R. Paris,
  W.~Parker, Z.~Patrick, A.~Pele, S.~Penn, M.~Phelps, M.~Pickenpack, V.~Pierro,
  I.~Pinto, J.~Poeld, M.~Principe, L.~Prokhorov, O.~Puncken, V.~Quetschke,
  E.~A. Quintero, F.~J. Raab, H.~Radkins, P.~Raffai, C.~R. Ramet, C.~M. Reed,
  S.~Reid, D.~H. Reitze, N.~A. Robertson, J.~G. Rollins, V.~J. Roma, J.~H.
  Romie, S.~Rowan, K.~Ryan, T.~Sadecki, E.~J. Sanchez, V.~Sandberg,
  V.~Sannibale, R.~L. Savage, R.~M.~S. Schofield, B.~Schultz, P.~Schwinberg,
  D.~Sellers, A.~Sevigny, D.~A. Shaddock, Z.~Shao, B.~Shapiro, P.~Shawhan,
  D.~H. Shoemaker, D.~Sigg, B.~J.~J. Slagmolen, J.~R. Smith, M.~R. Smith, N.~D.
  Smith-Lefebvre, B.~Sorazu, A.~Staley, A.~J. Stein, A.~Stochino, K.~A. Strain,
  R.~Taylor, M.~Thomas, P.~Thomas, K.~A. Thorne, E.~Thrane, C.~I. Torrie,
  G.~Traylor, G.~Vajente, G.~Valdes, A.~A. van Veggel, M.~Vargas, A.~Vecchio,
  P.~J. Veitch, K.~Venkateswara, T.~Vo, C.~Vorvick, S.~J. Waldman, M.~Walker,
  R.~L. Ward, J.~Warner, B.~Weaver, R.~Weiss, T.~Welborn, P.~We\ss{}els,
  C.~Wilkinson, P.~A. Willems, L.~Williams, B.~Willke, L.~Winkelmann, C.~C.
  Wipf, J.~Worden, G.~Wu, H.~Yamamoto, C.~C. Yancey, H.~Yu, L.~Zhang, M.~E.
  Zucker, and J.~Zweizig, \enquote{Sensitivity of the {Advanced LIGO} detectors
  at the beginning of gravitational wave astronomy,} Phys. Rev. D \textbf{93},
  112004 (2016).

\bibitem{Arai:2009cqg}
K.~Arai, R.~Takahashi, D.~Tatsumi, K.~Izumi, Y.~Wakabayashi, H.~Ishizaki,
  M.~Fukushima, T.~Yamazaki, M.-K. Fujimoto, A.~Takamori, K.~Tsubono,
  R.~DeSalvo, A.~Bertolini, S.~M{\'a}rka, V.~Sannibale, the TAMA~Collaboration,
  T.~Uchiyama, O.~Miyakawa, S.~Miyoki, K.~Agatsuma, T.~Saito, M.~Ohashi,
  K.~Kuroda, I.~Nakatani, S.~Telada, K.~Yamamoto, T.~Tomaru, T.~Suzuki,
  T.~Haruyama, N.~Sato, A.~Yamamoto, T.~Shintomi, the CLIO~Collaboration, and
  T.~L. Collaboration, \enquote{{Status of Japanese gravitational wave
  detectors},} Class. Quantum Grav. \textbf{26}, 204020 (2009).

\bibitem{Mullavey:2012oex}
A.~J. Mullavey, B.~J.~J. Slagmolen, J.~Miller, M.~Evans, P.~Fritschel, D.~Sigg,
  S.~J. Waldman, D.~A. Shaddock, and D.~E. McClelland, \enquote{Arm-length
  stabilisation for interferometric gravitational-wave detectors using
  frequency-doubled auxiliary lasers,} Opt. Express \textbf{20}, 81--89 (2012).

\bibitem{Izumi:2012josaa}
K.~Izumi, K.~Arai, B.~Barr, J.~Betzwieser, A.~Brooks, K.~Dahl, S.~Doravari,
  J.~C. Driggers, W.~Z. Korth, H.~Miao, J.~Rollins, S.~Vass, D.~Yeaton-Massey,
  and R.~X. Adhikari, \enquote{Multicolor cavity metrology,} J. Opt. Soc. Am. A
  \textbf{29}, 2092--2103 (2012).

\bibitem{Aso:2004pla}
Y.~Aso, M.~Ando, K.~Kawabe, S.~Otsuka, and K.~Tsubono, \enquote{{Stabilization
  of a Fabry-Perot interferometer using a suspension-point interferometer},}
  Phys. Lett. A \textbf{327}, 1 -- 8 (2004).

\bibitem{Shaddock:2007ol}
D.~A. Shaddock, \enquote{Digitally enhanced heterodyne interferometry,} Opt.
  Lett. \textbf{32}, 3355--3357 (2007).

\bibitem{Camp:1995ol}
J.~Camp, L.~Sievers, R.~Bork, and J.~Heefner, \enquote{Guided lock acquisition
  in a suspended {Fabry--Perot} cavity,} Opt. Lett. \textbf{20}, 2463--2465
  (1995).

\bibitem{Bersanetti.PhD}
D.~Bersanetti, \enquote{{Development of a new lock acquisition strategy for the
  arm cavities of Advanced VIRGO},} Ph.D. thesis, University of Genova (2016).

\bibitem{Takahashi:2008cqg}
R.~Takahashi, K.~Arai, D.~Tatsumi, M.~Fukushima, T.~Yamazaki, M.-K. Fujimoto,
  K.~Agatsuma, Y.~Arase, N.~Nakagawa, A.~Takamori, K.~Tsubono, R.~DeSalvo,
  A.~Bertolini, S.~M{\'a}rka, V.~Sannibale, and the TAMA~Collaboration,
  \enquote{Operational status of {TAMA300} with the seismic attenuation system
  {(SAS)},} Class. Quantum Grav. \textbf{25}, 114036 (2008).

\bibitem{Nagano:2003RSI}
S.~Nagano, S.~Kawamura, M.~Ando, R.~Takahashi, K.~Arai, M.~Musha, S.~Telada,
  M.-K. Fujimoto, M.~Fukushima, Y.~Kozai, S.~Miyama, A.~Ueda, K.~Waseda,
  T.~Yamazaki, H.~Ishizuka, K.~Kuroda, S.~Matsumura, O.~Miyakawa, S.~Miyoki,
  M.~Ohashi, S.~Sato, D.~Tatsumi, T.~Tomaru, T.~Uchiyama, K.~Kawabe, N.~Ohishi,
  S.~Otsuka, A.~Sekiya, A.~Takamori, S.~Taniguchi, K.~Tochikubo, K.~Tsubono,
  K.~Ueda, K.~Yamamoto, N.~Mio, S.~Moriwaki, G.~Horikoshi, N.~Kamikubota,
  Y.~Ogawa, Y.~Saito, T.~Suzuki, K.~Nakagawa, K.-I. Ueda, A.~Araya, N.~Kanda,
  N.~Kawashima, E.~Mizuno, M.~A. Barton, N.~Tsuda, N.~Matsuda, T.~Nakamura,
  M.~Sasaki, M.~Shibata, H.~Tagoshi, T.~Tanaka, K.-I. Nakao, K.-I. Oohara,
  Y.~Kojima, T.~Futamase, and H.~Asada, \enquote{{Development of a multistage
  laser frequency stabilization for an interferometric gravitational-wave
  detector},} Review of Scientific Instruments \textbf{74}, 4176 (2003).

\bibitem{Drever:1983apb}
R.~W.~P. Drever, J.~L. Hall, F.~V. Kowalski, J.~Hough, G.~M. Ford, A.~J.
  Munley, and H.~Ward, \enquote{Laser phase and frequency stabilization using
  an optical resonator,} Appl. Phys. B \textbf{31}, 97--105 (1983).

\bibitem{Izumi:2014ol}
K.~Izumi, D.~Sigg, and L.~Barsotti, \enquote{Self-amplified lock of an
  ultra-narrow linewidth optical cavity,} Opt. Lett. \textbf{39}, 5285--5288
  (2014).

\bibitem{Rakhmanov:2001ao}
M.~Rakhmanov, \enquote{Doppler-induced dynamics of fields in {Fabry--Perot}
  cavities with suspended mirrors,} Appl. Opt. \textbf{40}, 1942--1949 (2001).

\bibitem{e2e:2006}
H.~Yamamoto, M.~Barton, B.~Bhawal, M.~Evans, and S.~Yoshida,
  \enquote{Simulation tools for future interferometers,} J. Phys. Conf. Ser
  \textbf{32}, 398 (2006).

\end{thebibliography}

\end{document}